\documentclass{emulateapj}
\usepackage{graphicx,color}
\usepackage{amssymb,amsmath}
\usepackage[colorlinks,hyperfootnotes=false,citecolor=blue,linkcolor=blue]{hyperref}

\newcommand{\jcap}{JCAP}
\newcommand\Tstrut{\rule{0pt}{2.2ex}}         
\newcommand\Bstrut{\rule[-0.9ex]{0pt}{0pt}}   
\newcommand\Mstrut{\rule[-0.0ex]{0pt}{0pt}}

\begin{document}
\shorttitle{Strong Lensing Analysis of PLCK G287.0+32.9}
\shortauthors{Zitrin et al.}

\slugcomment{Submitted to the Astrophysical Journal Letters}

\title{A Very Large ($\theta_{E}\gtrsim40\arcsec$) Strong Gravitational Lens Selected with the Sunyaev-Zel'dovich Effect: PLCK G287.0+32.9 ($\lowercase{z} = 0.38$)}
\author{Adi Zitrin\altaffilmark{1,*}, Stella Seitz\altaffilmark{2,3}, Anna Monna\altaffilmark{2,3}, Anton M. Koekemoer\altaffilmark{4}, Mario Nonino\altaffilmark{5}, Daniel Gruen\altaffilmark{6,7,$\dagger$}, Italo Balestra\altaffilmark{2,5}, Marisa Girardi\altaffilmark{8,5}, Johannes Koppenhoefer\altaffilmark{2,3}, Amata Mercurio\altaffilmark{9}}
\altaffiltext{1}{Physics Department, Ben-Gurion University of the Negev, P.O. Box 653, Be'��er-Sheva 8410501, Israel\\ * adizitrin@gmail.com}
\altaffiltext{2}{University Observatory Munich, Scheinerstrasse 1, D-81679, Munich, Germany}
\altaffiltext{3}{Max Planck Institute for Extraterrestrial Physics, Giessenbachstrasse, D-85741 Garching, Germany}
\altaffiltext{4}{Space Telescope Science Institute, 3700 San Martin Drive, Baltimore, MD 21218, USA}
\altaffiltext{5}{INAF - Osservatorio Astronomico di Trieste, via G. B. Tiepolo 11, I-34131, Trieste, Italy}
\altaffiltext{6}{SLAC National Accelerator Laboratory, Menlo Park, CA 94025, USA}
\altaffiltext{7}{KIPAC, Physics Department, Stanford University, Stanford, CA 94305, USA}
\altaffiltext{8}{Dipartimento di Fisica, Universit\'a degli Studi di Trieste, Via Tiepolo 11, I-34143 Trieste, Italy}
\altaffiltext{9}{INAF - Osservatorio Astronomico di Capodimonte, Via Moiariello 16 I-80131 Napoli, Italy}
\altaffiltext{$\dagger$}{Einstein Fellow}

\begin{abstract}
Since galaxy clusters sit at the high-end of the mass function, the number of galaxy clusters both massive and concentrated enough to yield particularly large Einstein radii poses useful constraints on cosmological and structure formation models. To date, less than a handful of clusters are known to have Einstein radii exceeding $\sim40\arcsec$ (for a source at $z_{s}\simeq2$, nominally). Here, we report an addition to that list of the Sunyaev-Zel'dovich (SZ) selected cluster, PLCK G287.0+32.9 ($z=0.38$), the second-highest SZ-mass ($M_{500}$) cluster from the Planck catalog. We present the first strong lensing analysis of the cluster, identifying 20 sets of multiply-imaged galaxies and candidates in new \emph{Hubble Space Telescope} data, including a long, $l\sim22\arcsec$ giant arc, as well as a quadruply-imaged, apparently bright (magnified to J$_{F110W}=$25.3 AB), likely high-redshift dropout galaxy at $z_{phot}=6.90$ [6.13--8.43] (95\% C.I.). Our analysis reveals a very large critical area (1.55 arcmin$^{2}$, $z_{s}\simeq2$), corresponding to an effective Einstein radius of $\theta_{E}\sim42\arcsec$. The model suggests the critical area will expand to 2.58 arcmin$^{2}$ ($\theta_{E}\sim54\arcsec$) for sources at $z_{s}\sim10$. Our work adds to recent efforts to model very massive clusters towards the launch of the James Webb Space Telescope, in order to identify the most useful cosmic lenses for studying the early Universe. Spectroscopic redshifts for the multiply-imaged galaxies and additional HST data will be necessary for refining the lens model and verifying the nature of the $z\sim7$ dropout. \vspace{0.05cm}
\end{abstract}


\section{Introduction}\label{intro}
Galaxy clusters are the most massive gravitationally-bound objects in the Universe, sitting at the high end of the mass function. Their total masses and mass profiles can be inferred in various complementary ways, including from cluster galaxy kinematics \citep{DiaferioGeller1997}, weak lensing (WL) \citep{BartelmannSchneider2001}, X-ray measurements, \citep{Sarazin1986}, or the Sunyaev-Zel'dovich effect \citep{SunyaevZeldovich1972}. 

The abundance of the most massive clusters is important for scaling the mass function and probing cosmological models. N-body numerical simulations predict a universal mass profile form for virialized dark matter (DM) halos (e.g. \citealt{Navarro1996}),  with increasing concentrations towards lower redshifts (as structure evolves) and for smaller masses \citep{Duffy2008}. Generally, the Einstein radius of a lens increases with its overall mass (explicitly, with the inner projected mass density), and with concentration \citep{SadehRephaeli2008}. Given the shape of the concentration-mass relation, more massive, highly concentrated clusters become rarer. Assuming a mass function and a cosmological model, predictions for the distribution of Einstein radii in the Universe can be made \citep{OguriBlandford2009,Waizmann2012JeanClaude0717,Redlich2014}. Indeed, more factors can play an important role in determining the actual critical area size, such as the effective ellipticity and distribution of projected substructure \citep{Redlich2012MergerRE}, or the halo's triaxiality. These factors can be accounted for statistically so that the comparison of the abundance of large lenses to $\Lambda$CDM predictions remains interesting. 

To date, less than a handful of clusters are known to have critical areas exceeding 1.4 arcmin$^{2}$  for $z_{s}=2$, i.e. an effective Einstein radius (defined as $\sqrt{A/\pi}$, with $A$ being the critical area) exceeding 40\arcsec (and only few others are known with such large critical area even for high-redshift sources). The largest strong lens currently known, with $\theta_{E}\sim55\arcsec$ \citep[][]{Zitrin2009_macs0717}, is MACS J0717.5+3745, a merging, massive \citep{Medezinski2013M0717} X-ray selected \citep{Ebeling2010FinalMACS} galaxy cluster at $z=0.55$. In fact, most massive clusters with known large Einstein radii have been chosen for follow-up based on optical classification (e.g. richness and luminosity), or X-ray measurements \citep{Ebeling2010FinalMACS}. However, with the recent abundance of SZ data and related cluster catalogs, such as those from Planck \citep{Planck2011ClustersV1}, the South Pole Telescope (SPT; \citealt{SPT2010ClusterCat}) and the Atacama Cosmology Telescope (ACT, \citealt{Hasselfield2013ACTclusters}), more potentially massive galaxy clusters have been detected.

Here, we present the first strong-lensing (SL) analysis of the Planck galaxy cluster PLCK G287.0+32.9 (PLCKG287 hereafter), which has the second highest SZ-inferred mass in the Planck catalog\footnote{PSZ1; this is Mass\_YZ\_500, an SZ mass proxy based on X-ray calibration of scaling relations.} ($M_{SZ,500}=13.89^{+0.53}_{-0.54}\times10^{14}~M_{\odot}$, \citealt{Planck2015ClusterCat}). 

This Letter is organized as follows: In \S \ref{obs} we summarize the observations and data reduction. In \S \ref{lensmodel} we review the SL modeling of the cluster. In \S \ref{discussion} we discuss the analysis results and their potential implications, and in \S \ref{summary} we conclude the work. Throughout we use a standard $\Lambda$CDM cosmology with $\Omega_{\rm m0}=0.3$, $\Omega_{\Lambda 0}=0.7$, $H_{0}=100$ $h$ km s$^{-1}$Mpc$^{-1}$, $h=0.7$. Magnitudes are given using the AB convention. 1\arcsec\ equals 5.21 kpc at the redshift of the cluster. Unless noted otherwise, errors are $1\sigma$.

\begin{figure*}
 \begin{center}
   \includegraphics[width=177mm,trim=0cm 0cm 0cm 0cm,clip]{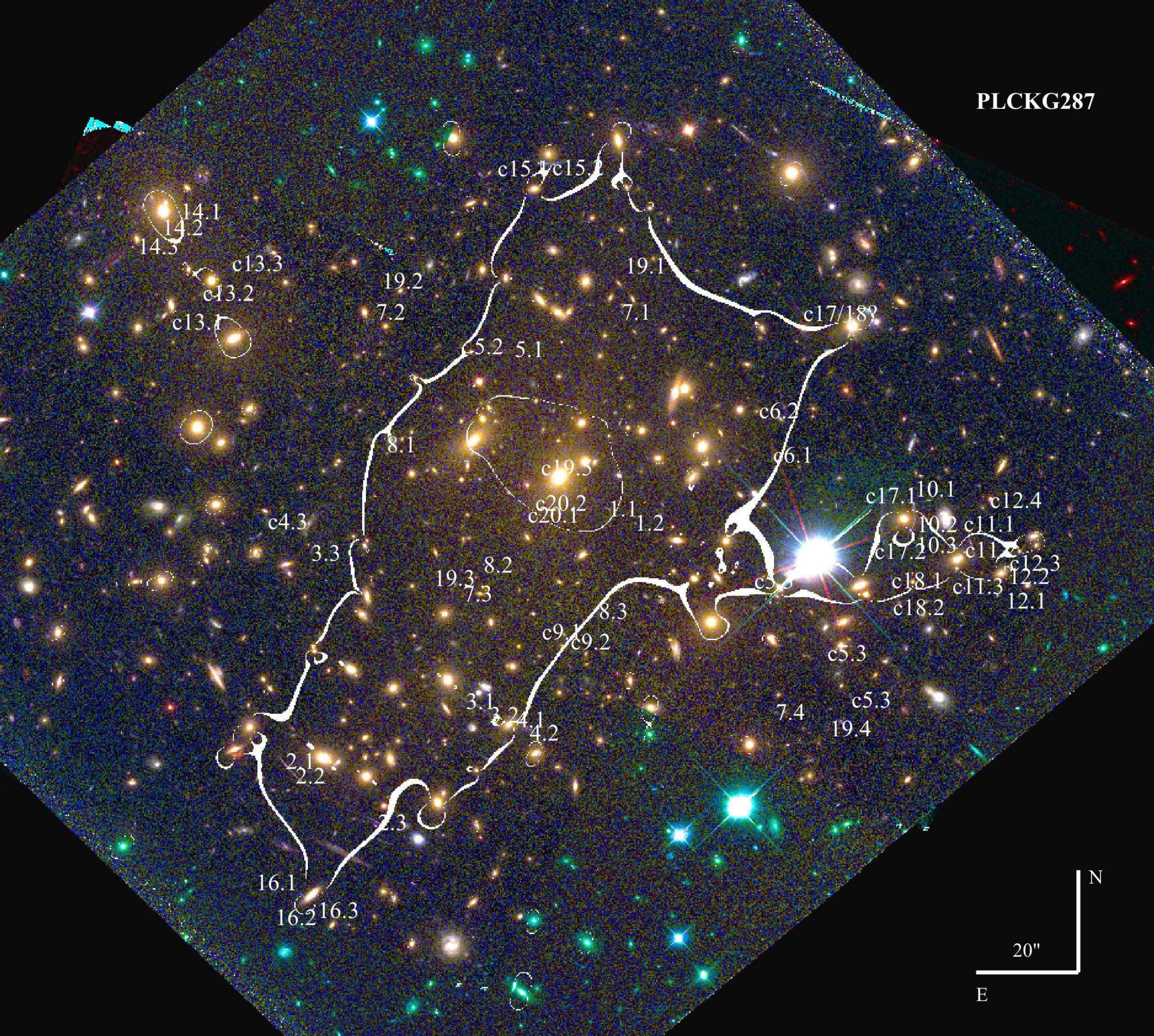}
 \end{center}
\caption{Galaxy cluster PLCKG287. The critical curves from our model are marked in white for a source at $z_{phot}\simeq3.4$, enclosing an area with an effective Einstein radius of $\theta_{E}=49\pm5\arcsec$. The multiple images we identify are numbered and marked on the image (``c" stands for ``candidate''). Image is constructed using R$=$F110W, G$=$F814W+F606W, B$=$F475W.}\vspace{0.05cm}
\label{fig1}
\end{figure*}

\section{Target, Data and Observations}\label{obs}
PLCGK287 was discovered in early SZ observations by the
Planck telescope \citep{Planck2011ClustersV1}. Follow-up X-ray observations with
XMM-Newton yield a temperature of $T_{X} = 12.86 \pm0.42$ keV and mass of $M_{500}=15.72\pm0.27\times10^{14}~h_{70}^{-1}M_{\odot}$ \citep{Planck2011XMM}.

The cluster was found to show also a non-thermal radio emission characterized by a double relic, indicating the system has undergone a recent major merger \citep{Bagchi2011PLCKG287,Bonafede2014}. The projected separation of the double radio relic is  $\sim4$ Mpc, the largest known at comparable redshifts. From the XMM-Newton data a separation of $\sim400$ kpc is measured between the cluster BCG and the X-ray emission peak, providing additional evidence for merger activity.

PLCKG287 was previously imaged with the WFI on the 2.2 m MPG/ESO telescope in VRI filters. In a systematic WL study of SZ-selected clusters,  \citet[][see also K. Finner et al., in preparation]{Gruen2014} confirmed the large mass estimates from SZ and X-ray and found the system to be the most massive among their sample, with $M_{WL,500} = 19.5^{+3.3}_{-3.2}\times10^{14}~h_{70}^{-1}M_{\odot}$. In these optical data several arc-like features were identified, spread out over a large field around the cluster core, with a radius exceeding an arcminute. 

Following these findings we obtained \emph{Hubble Space Telescope} (HST) data in Cycle 23 to perform a dedicated SL analysis of this cluster (PI: Seitz, program ID 14165).  PLCKG287 was observed on 2016 August 3 in the optical, ACS bands F475W, F606W and F814W, with total exposure times of 1, 1, and 2 orbits, respectively. The cluster was also observed on 2016 May 18 with the WFC3/IR F110W filter, with 4 exposures per orbit over 4 orbits, centered on two locations, and adopting a half-pixel dither pattern. Data reduction was performed using advanced drizzle techniques \citep{Koekemoer2011}, including CR rejection, full astrometric alignment, weighting by the inverse variance, and drizzling, to produce
the final set of mosaics.

We ran SExtractor \citep{BertinArnouts1996Sextractor} in dual-image mode to obtain the photometry of objects in the cluster field, using the F814W image as reference. Red-sequence cluster members, needed for our modeling, are chosen by a color-magnitude diagram, using the F814W and F606W filters, down to 23 AB. We used the photometric catalogs as input and ran the Bayesian Photometric Redshift program (BPZ; \citealt{Benitez2000}), to derive redshift estimates and help in the identification of multiple images\footnote{We also generated a version with the F110W image as reference}.

We had also observed PLCKG287 with VIMOS\footnote{Programme 094.A-0529} on the VLT/UT3 (\citealt{LeFevreVLT}) in service mode on four nights in February-March 2015. We used the Medium Resolution red grism and the GG475 filter, covering the $\approx 5000-10,000\AA$ spectral range, with a resolution of $\approx 600$. A total of twenty 1135 sec exposures were obtained. These data have been reduced with a mix of custom pipeline and IRAF tasks ({\it apall, onedspec}). IRAF {\it rvsao} (\citealt{KurtzMink1998RVSAO}) was used to estimate the redshifts. From $\ge 220$ member galaxies we estimate $z=0.380$ for the cluster redshift (updating the $z=0.39$ Planck estimate).

We cross-check our red-sequence selection with the VIMOS data to maximize the number of verified members included, and remove red non-members. Out of the 248 red-sequence galaxies we consider members for the modeling, 35 within the HST FOV are spectroscopically confirmed. Additional objects suspected as interlopers in a visual inspection were removed. 

\section{Lens Model} \label{lensmodel}
We use the light-traces-mass (LTM) method by \citet[][see also \citealt{Broadhurst2005a,Zitrin2014CLASH25}]{Zitrin2009_cl0024} for the SL analysis of PLCKG287. \\
The photometry of red-sequence cluster galaxies (\S \ref{obs}) is the starting point for the model.  A power-law surface mass-density distribution is assigned to each galaxy, scaled in proportion to its luminosity (we use here the F606W magnitudes as reference). For the BCGs we typically assign $\emph{elliptical}$ power-law surface mass-density distributions with a core. The power-law exponent is the same for all galaxies and is the first free parameter of the model. The resulting map from the superposition of these galaxies is then smoothed with a Gaussian kernel, whose size is the second free parameter. The smoothed map constitutes the DM component of our model. The two components are then added with a relative weight, which is also iterated for, and scaled with an overall normalization, which is the fourth free parameter. To allow for further flexibility we also add a two-parameter external shear. The total number of free parameters principally is thus six. In addition, we typically also allow for the weight (and possibly, core radius and ellipticity) of the brightest galaxies to be freely optimized in the minimization procedure.

Using preliminary LTM models, a method that has been shown to excel in predicting the appearance of multiply-imaged galaxies \citep[e.g.][]{Broadhurst2005a,Zitrin2009_cl0024,Zitrin2009_macs0717}, we iteratively identify 20 sets of multiple images and candidates in the HST data (\S \ref{obs}; see Figure \ref{fig3}). We set their redshift to the corresponding, best-value photometric redshifts, and use them as constraints for constructing the final model presented here. We only fix the redshifts for systems with relatively secure estimates (dropouts, or those where the photo-$z$ agrees well among the different multiple images). The redshifts of the other systems were left free to be optimized in the minimization, as indicated in Table 1 listing the multiple images. 

For the two central BCGs, we assign ellipticity values measured by SExtractor. We leave the weight of the two central BCGs, and two other bright galaxies, to be optimized in the minimization procedure.

\begin{figure}
 \begin{center}
   \includegraphics[width=92mm,trim=2.5cm 0cm 2cm 3cm,clip]{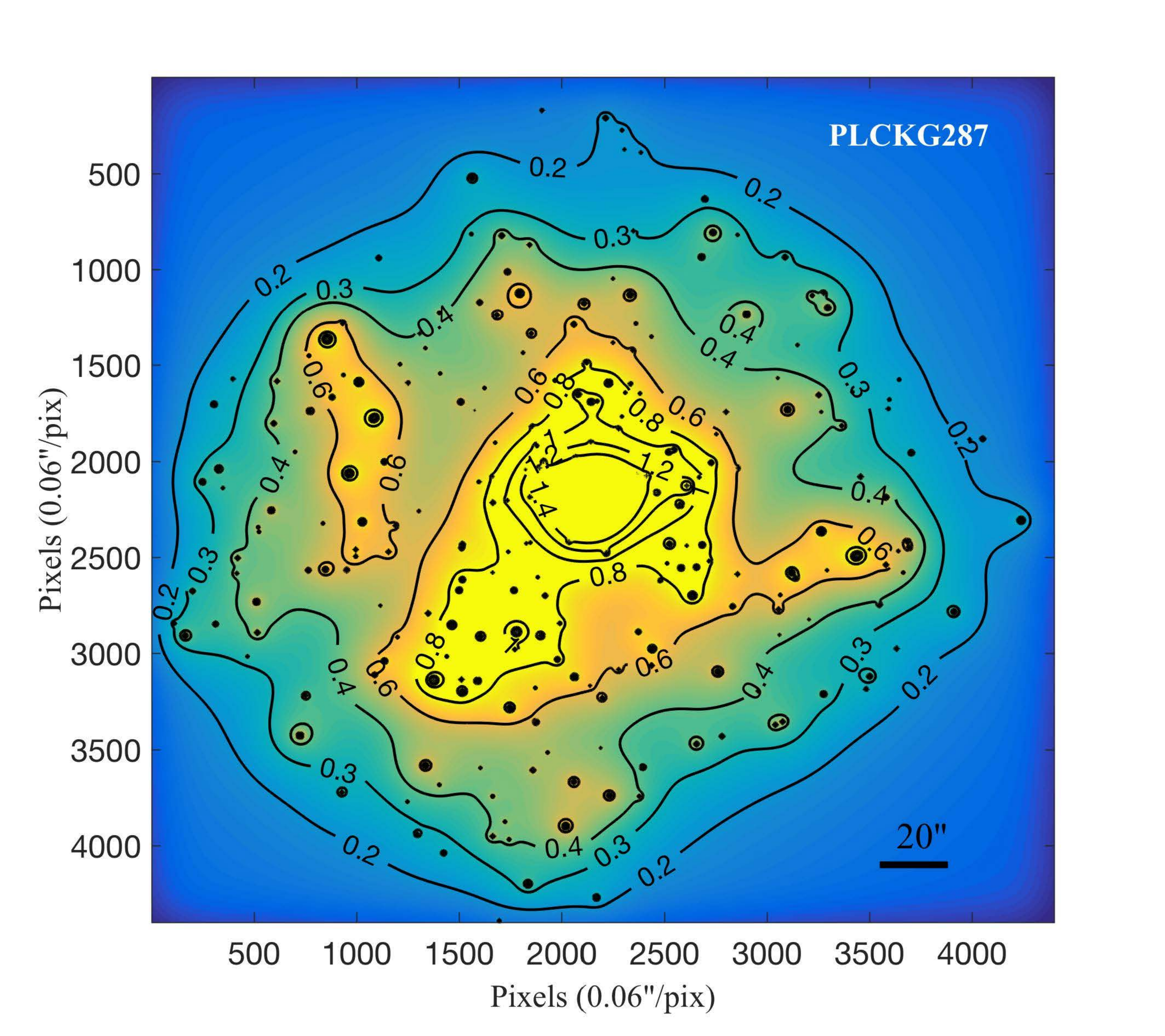}
   \includegraphics[width=92mm,trim=2.5cm 0cm 2cm 3cm,clip]{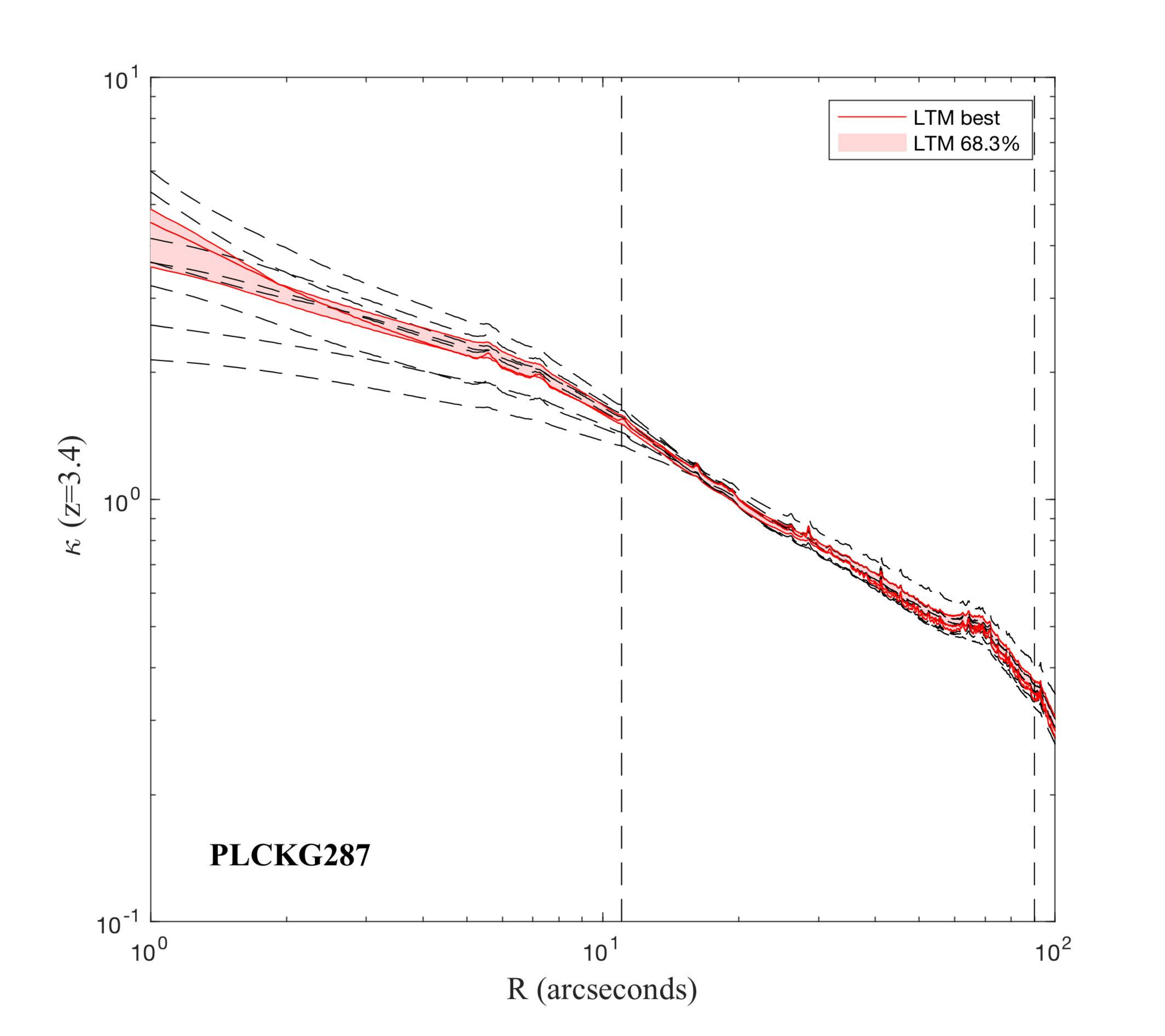}
 \end{center}
\caption{\emph{Top:} map of $\kappa$, the projected surface mass density in units of the critical density for lensing, scaled to the redshift of system 7, $z_{s}\simeq3.4$. \emph{Bottom:} the corresponding radially-averaged mass density profile. The profile slope in the range [1,84] arcseconds ($\sim440$ kpc, about twice the Einstein radius), is $d\log \kappa / d\log \theta \simeq-0.55$, similar to other well known lensing clusters (Fig. 7 in \citealt{Zitrin2014CLASH25}).  Vertical dashed lines mark the area in which there are multiple-image constraints. The black, dashed profile lines demonstrate the range spanned by models run with different choices of fixed source redshifts.}\vspace{0.05cm}
\label{fig2}
\end{figure}

\begin{figure}
 \begin{center}
     \includegraphics[width=88mm,trim=0cm 0cm 0cm 0cm,clip]{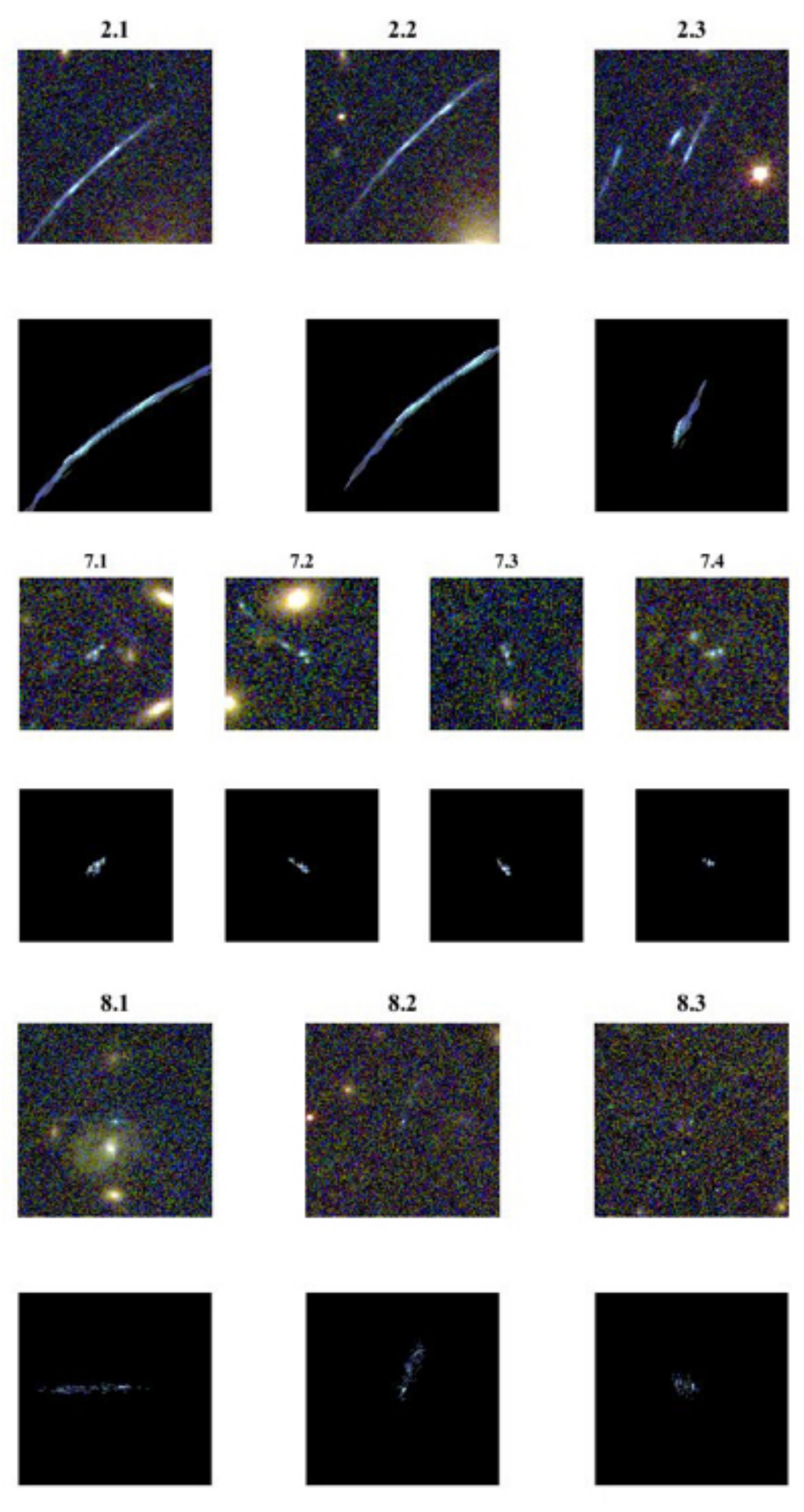}
 \end{center}
\caption{Reproduction of multiple images by our model. In each case we delens one of the images (2.2, 7.1, and 8.2, explicitly) to the source plane, and back to the image plane to compare to the other images of that system. As can be seen, the orientation and internal details of the predicted images (bottom rows) resemble those of the real images identified in the predicted location in the data (upper rows), supporting the identification.}\vspace{0.05cm}
\label{fig3}
\end{figure}

The minimization of the model is performed with a several thousand Monte Carlo Markov Chain (MCMC) steps, through a $\chi^{2}$ criterion quantifying the reproduction of the multiple image positions (adopting a positional uncertainty of 0.5\arcsec, or 1.4\arcsec when calculating the errors). The final model has an image reproduction \emph{rms} of 1.9\arcsec. The resulting critical curves ($z_{s}=3.4$) are seen in Fig. \ref{fig1} along with the multiply-imaged galaxies. The best-model mass map and profile are seen in Fig. \ref{fig2}.

It should be noted that our results are based only on 4-band photometric redshifts for the multiple images, and thus may vary strongly depending on the true redshift of the sources. In addition, there is some ambiguity with respect to few potential multiple images (Fig. \ref{fig1}, Table 1; for example, which candidates are the true, third counter images of systems 5, 12, or 15). It is also at present unclear what is the nature of the $22\arcsec$ long giant arc (systems c17 and c18), whether it is multiply imaged and whether there is an additional counter image north of the arc. Similarly, if candidate system 13 is a multiply-imaged galaxy, this means that the eastern ``arm" seen in the surface density map in Fig. 2, creates a local critical curve larger that the simple LTM assumption yields. On the other hand, if systems 10 and 11 have a notably higher redshift than our current estimates imply, it might mean the western arm is in fact smaller and consists only of local critical curves around the respective galaxies.\\
Additional uncertainty arises from the bright galaxy at [RA,DEC]$=$[11:50:46.747,-28:03:56.560] that was chosen by our red sequence criteria, but the VIMOS data suggests it is not officially a cluster member (it has a $>$10,000 km/s lower velocity than the cluster average). The galaxy seems to contribute to some extent to the lensing signal, though, and so we leave it in the modeling but with a reduced weight.  The lack of obvious multiple-image systems around that northern tip suggests this choice is reasonable.

We did not use candidate images as constraints, and, in most ambiguities listed above we generally favored the conservative choice, so that the actual critical area may, if anything, be larger than our estimate in \S \ref{discussion}.

\section{Results and Discussion} \label{discussion}
Our SL analysis reveals a very big lens, with a critical area of 1.55 arcmin$^{2}$, for $z_{s}\simeq2$, corresponding to an effective Einstein radius of $\theta_{E}\simeq 42\pm4\arcsec$. The mass enclosed within these curves is $3.1\pm0.5\times10^{14}~M_{\odot}$.  For $z_{s}\simeq3.4$, for example, the critical curves expand further as expected, reaching an effective Einstein radius of $\theta_{E}\sim49\arcsec$, and for $z_{s}\sim10$, our model suggests they would reach $\theta_{E}\sim54\arcsec$.

To examine the effect of lack of accurate redshifts we ran four other models with significantly different combinations of fixed redshifts. The Einstein radii estimates between the different models is within 10\% of each other, and the enclosed masses agree to 15\%. The mass profile and correspondingly, predicted redshifts, are more sensitive to the exact redshifts initially adopted, and typically agree to $2\sigma-3\sigma$.

Few clusters are known to have Einstein radii above 40\arcsec (for typical source redshifts of $z_{s}\sim2$). These include the largest known gravitational lens, MACS J0717.5+3745, included also in the Cluster Lensing And Supernova with Hubble (CLASH; \citealt{PostmanCLASHoverview}) and Hubble Frontier Fields (HFF; \citealt{Lotz2016HFF}) programs, with $\theta_{E}\sim55\arcsec$ \citep{Zitrin2009_macs0717}; Abell 1689 ($\theta_{E}\sim45\arcsec$, \citealt{Broadhurst2005a}); and the HFF cluster Abell 370 ($\sim40\arcsec$, \citealt[][]{Richard2010A370}). Worth mentioning is also RCS2 J232727.6-020437, for which \citet{Sharon2015RSC2Big} find $\theta_{E}\simeq40\arcsec$ for $z_{s}\simeq3$  (but smaller -- i.e. 26\arcsec -- as expected, for another source at $z\simeq1.4$).

The distribution of Einstein radii, and in particular the high-end of this distribution, is important to characterize, as the largest and most massive lenses help to probe cosmological models and structure formation scenarios \citep{OguriBlandford2009,Waizmann2012JeanClaude0717,Redlich2014}. The high concentrations and large Einstein radii found for several massive clusters have been previously claimed to challenge $\Lambda$CDM \citep{Broadhurst2008,BroadhurstBarkana2008}, although updated analyses (and account of projection biases) have alleviated this tension \citep{Merten2015CLASH,Umetsu2016CLASH}. In a similar fashion, the amount of substructure within massive clusters can also be compared to numerical simulations and expectations from $\Lambda$CDM  \citep{Jauzac2016A2744Substructure,Schwinn2016A2744}. Given its extreme properties, PLCKG287 is another useful laboratory for similar studies.

Most of the impressive lenses known to date, including the $\theta_{E}>40\arcsec$ lenses mentioned above as well as those observed by the CLASH and HFF programs, and that were found to be magnifying many high redshift lensed galaxies \citep[e.g.][and references therein]{Bradley2013highz,Kawamata2016HFF}, were selected for HST follow-up mostly due to their optical (e.g. richness), or X-ray signals. X-ray selection in particular has proved to be an excellent probe for locating merging clusters with large critical curves \citep{Ebeling2010FinalMACS,Zitrin2014CLASH25}, as mergers tend to boost the critical area. Here, we concentrate on an SZ-selected cluster, which has the second highest $M_{SZ}$ estimate in the Planck cluster catalog. In addition to our observations, PLCKG287 will be observed in the ongoing Reionization Lensing Cluster Survey (RELICS, PI: Coe), an HST program to observe a large set of mainly SZ-mass selected clusters, designed to find high-redshift, bright dropout galaxies in the reionization epoch.

Indeed, recent SZ surveys have expanded our massive-cluster sample. One such notable example, is ACT-CL J0102-4915, the "El Gordo" cluster, a high-redshift ($z=0.87$), and likely the most massive, hottest, most X-ray luminous and brightest SZ effect cluster known at $z>0.6$ \citep{Menanteau2012Gordo,Jee2014Gordo}. \citet{Zitrin2013Gordo} performed the first SL analysis of this cluster, revealing a large lens with a critical area exceeding $\simeq1.4$ arcmin$^{2}$ for high-redshift sources ($z_{s}\gtrsim4$). This shows the power of SZ massive-cluster selection, especially at higher redshifts, much due to the fact that the SZ signal is not redshift dependent (nonetheless, X-ray and optical / near-infrared observations have also revealed high-redshift clusters, including out to $z>1$ and $z>2$, respectively; \citealt{Rosati2009,Strazzullo2016z2cluster}).

One interesting question is whether there is a preferred selection that leads to larger or ``better" strong lenses, particularly for high-$z$ applications. Scaling relations between X-ray, SZ, luminosity, richness, and lensing masses are well established, and characterized with increasing precision \citep[e.g.][]{Rozo2014Scaling}.  However, high total mass is not sufficient to guarantee a large critical area, which is dependent on the exact central projected mass density distribution \citep{Redlich2012MergerRE}. In this relation, efforts have been made in recent years to detect clusters with large critical areas more directly, based on the luminosity distribution of red cluster member galaxies, and calibrated with well-studies lensing clusters \citep{Zitrin2012UniversalRE,Wong2012OptLenses}. It will be interesting to compare these selections to other probes (richness, SZ, X-ray) once relevant HST data become available. 

\begin{figure}
 \begin{center}
  \includegraphics[width=88mm,trim=0.75cm 3cm 0.2cm 0.5cm,clip]{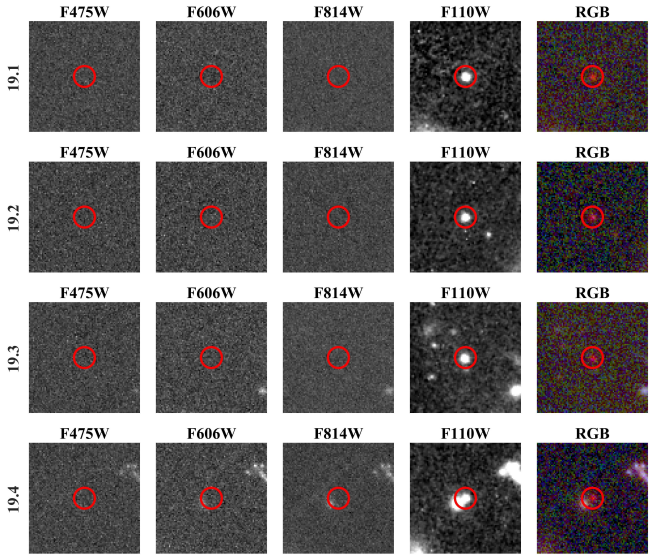}
 \end{center}
\caption{Quadruply-lensed, dropout high-redshift candidate (System 19). All four images of the dropout galaxy are detected only in the F110W band. Squares are 5\arcsec on a side. Red circles' radius is $\simeq0.5$\arcsec.}\vspace{0.1cm}
\label{fig4}
\end{figure}

Among the multiply-imaged galaxies we detect in PLCKG287, noteworthy is a likely high-redshift dropout, detected only in F110W. We identify four multiple images of this galaxy (Fig. \ref{fig4}). An additional, demagnified fifth one is predicted next to the BCG core, and we list here one potential candidate (system 19, Table 1). The photometric redshifts for all four magnified images of the galaxy agree well, $z_{phot}=6.9$ [6.13--8.43] (95\% C.I.), and corroborated by our lens model: the system symmetry follows that of system 7 at $z\simeq3.4$, but at a larger radius (and requires a higher lensing distance ratio) so that the higher-redshift nature of this object is also supported geometrically from lensing. 

Due to its relative brightness (observed apparent magnitudes of J$_{F110W}$=25.3-25.5 AB for the four images), this galaxy presents a promising case for the challenging spectroscopic follow-up of high-redshift sources. In addition to its brightness, the SNR of such observations can be significantly increased by observing the four magnified images simultaneously. Using the magnification factors implied by our model, we obtain the source intrinsic (i.e. demagnified) apparent magnitude is J$_{F110W}\sim26.7\pm0.4$ AB, and the intrinsic half-light radius is $0.4\pm0.14$ kpc, where the errors represent the range from the four images and their magnification values. In future iterations, the source's relative magnifications can also be used as additional constraints for the model.

\section{summary}\label{summary}
Galaxy clusters constitute great cosmic telescopes. Extensive lensing surveys have taught us that essentially all massive clusters act as useful gravitational lenses, distorting, magnifying, and multiply-imaging objects behind them. This has allowed us to construct dozens of mass maps and enabled the continuous discovery of hundreds of magnified, high-redshift galaxies.  However, even with significant HST time devoted to recent lensing surveys, the data (and corresponding lensing models) currently available, sample only a relatively small part of all massive clusters. There are expected to be a dozen to several dozen massive clusters in the sky with $\theta_{E}>40\arcsec$ \citep{OguriBlandford2009,Zitrin2012UniversalRE}, and finding most massive lenses and the best cosmic telescopes among these is important for studying the cosmological model and structure formation and evolution, and for maximizing the chances to detect increasingly-fainter objects at higher redshifts. 

Here, we modeled the massive cluster PLCKG287, and identified 20 multiple-image families and candidates, including an apparently bright (J$_{F110W}=$25.3 AB), quadruply-imaged $z_{phot}\sim7$ galaxy. Our analysis reveals a large Einstein radius of $42\pm4\arcsec$ for $z_{s}\simeq2$, and $54\pm5\arcsec$ for $z_{s}\sim10$, adding to the short list of only few similarly large lenses. Our results are based solely on photometric redshifts, and so a refinement of the model is warranted when more data -- in particular spectroscopic redshifts --  become available (however given the redshift range already spanned by the multiply-imaged galaxies, we consider the Einstein radius robust). Interestingly, in contrast to most impressive lenses we are familiar with to date, that are X-ray or optically selected, and similar to the notable example of \emph{El Gordo}, this cluster is SZ-selected. It will be intriguing to see in the near future, if a certain selection (such as SZ, X-ray, luminosity, richness, or other, more direct lensing-based optical methods), is most efficient in locating the largest strong lenses and those best suited for studying the early Universe.  
\\
\section*{acknowledgments}
We thank the anonymous reviewer of this work for very useful comments. This work is based on observations made with the NASA/ESA Hubble Space Telescope. Support for program ID 14165 (PI: Seitz) was provided by NASA from the Space Telescope Science Institute (STScI), which is operated by the Association of Universities for Research in Astronomy, Inc. under NASA contract NAS 5-26555. S. Seitz thanks the DFG Transregio 33 ``The Dark Universe" and the DFG cluster of excellence``Origin and Structure of the Universe" for support. Support for DG was provided by NASA through the Einstein Fellowship Program, grant PF5-160138. AM and MN acknowledge support by PRIN-INAF 2014 1.05.01.94.02. IRAF is distributed by the National Optical Astronomy Observatories, which are operated by the Association of Universities for Research in Astronomy, Inc., under cooperative agreement with the National Science Foundation. IRAF {\it rvsao} was developed at the Smithsonian Astrophysical Observatory Telescope Data Center.

\begin{deluxetable*}{|lccccc|}
\tablecaption{\small{Multiple Images and Candidates} \label{multTable}}
\tablehead{
\colhead{Arc ID
} &
\colhead{R.A
} &
\colhead{DEC.
} &
\colhead{$z_{phot}$ [95\% C.I.]
} &
\colhead{$z_{model}$ [95\% C.I.]
} &
\colhead{Comments}
}
\startdata
1.1 & 11:50:49.448&-28:05:02.060 & 3.76 [3.29--4.23] &$=$3.75 & Radial image \Mstrut \\
1.2  &11:50:49.054 &-28:05:04.792 & 3.72 [3.26--4.18]  &  " & $"$ \Mstrut \\
\hline 
2.1 &11:5054.201 &-28:05:52.064   & 1.67 [1.41--1.93]  &  1.38 [1.30--1.60]  & Other iterations yield $z_{phot}\sim2.3$ \Tstrut \\
2.2 &11:50:54.060 &-28:05:54.105   & 1.82 [1.54--2.10]  &  " & " \Mstrut \\
2.3 &11:50:52.844 &-28:06:03.226   & 1.67 [1.41--1.93]   &  " & " \Bstrut \\
\hline
3.1 &11:50:51.549 &-28:05:39.538   & 2.81 [0.04--3.18] &   1.34 [1.21--1.46] & \nodata \Tstrut \\
3.2 &11:50:51.197 &-28:05:42.271   & 0.35 [0.09--3.12]  &  " & \nodata  \Mstrut \\
3.3 &11:50:53.830 &-28:05:10.648   & 0.93 [0.13--2.56]  &  " &  \nodata \Bstrut \\
\hline
4.1 &11:50:50.800 &-28:05:43.124   & 1.04 [0.84--1.24]  &  2.55 [2.14--2.92] & \nodata \Tstrut \\
4.2 &11:50:50.667 &-28:05:44.562   & 1.17  [0.96--1.41]  &  "& \nodata \Mstrut \\
c4.3  &11:50:54.469 &-28:05:04.469  & 0.97 [0.13--1.25]  &  " &\nodata  \Bstrut \\
\hline
5.1  &11:50:50.890 &-28:04:31.287  & 2.55 [0.07--2.93]  & 1.53 [1.31--1.89] & \nodata \Tstrut \\
5.2 &11:50:51.303 &-28:04:31.069   & 2.23 [0.18--2.81] & " & \nodata \Mstrut \\
c5.3 &11:50:47.248 &-28:05:16.134   & 1.86 [1.58--2.41] &  " & \nodata \Mstrut \\
c5.3 & 11:50:46.243 &  -28:05:30.772 & 3.14 [1.14--3.57] &  " &  Most likely according to the model \Mstrut \\
c5.3 &11:50:45.865 &-28:05:39.327   &  0.46 [0.13--3.31] &  " & \nodata \Bstrut \\
\hline
c6.1  &11:50:47.021 &-28:04:51.560  & 4.43 [3.90--4.96]  &  \nodata & \nodata  \Tstrut \\
c6.2  &11:50:47.232 &-28:04:42.851  & 4.51 [0.68--5.05]  &  \nodata  &\nodata   \Bstrut \\
\hline
7.1 &11:50:49.252 &-28:04:23.606   & 3.32 [2.90--3.74]  &  $=$3.4 & \nodata  \Tstrut \\ 
7.2 &11:50:52.875 &-28:04:23.606  & 3.47 [3.03--3.91]  &  " & \nodata  \Mstrut \\
7.3  &11:50:51.702 &-28:05:17.047 & 3.33 [2.91--3.79]  &  " & \nodata  \Mstrut \\
7.4  &11:50:46.982 &-28:05:41.706 & 3.52 [3.08--3.96] &  " & \nodata  \Bstrut \\
\hline
8.1 &11:50:52.743 &-28:04:48.339   & 2.44 [1.88 2.78]  &  0.99 [0.92--1.06] & \nodata \Tstrut \\
8.2  &11:50:51.290 &-28:05:12.952  & 3.04  [3.59--3.44] &  " & Other iterations yield $z_{phot}\sim2.4$ \Mstrut \\
8.3 &11:50:49.593 &-28:05:22.032 & 1.06 [0.45--1.66] &  " & "  \Bstrut \\
\hline
c9.1 &11:50:50.430 &-28:05:26.016  &  0.87 [0.59--2.78]& \nodata & \nodata \Tstrut \\
c9.2 &11:50:50.200 &-28:05:27.985  &  0.85 [0.22--4.06] & \nodata & Nearby parts yield  $z_{phot}\sim2.5$ \Bstrut \\
\hline
10.1  &11:50:44.942 &-28:05:02.578  & 0.98 [0.79--1.17]  &  0.89 [0.71--0.91] & \nodata \Tstrut \\
10.2 &11:50:44.929 &-28:05:04.072  & 0.97 [0.78--1.16]  &  " & \nodata \Mstrut \\
10.3  &11:50:44.993 &-28:05:06.556 & 1.02 [0.82--1.22]  &  " & \nodata \Bstrut \\
\hline
c11.1 &11:50:44.089 &-28:05:09.461  & 0.10 [0.00--2.54] &  \nodata & \nodata \Tstrut \\
c11.2 &11:50:44.063 &-28:05:10.642  & 0.09 [0.00--3.09]  &  \nodata & \nodata \Mstrut \\
c11.3 &11:50:44.202 &-28:05:17.428  & 0.09 [0.00--2.86]   &  \nodata& \nodata \Bstrut \\
\hline
12.1 &11:50:43.705 &-28:05:18.794 & 3.43 [3.00--3.86]  &  $=$3.4 & \nodata \Tstrut \\
12.2 &11:50:43.598 &-28:05:13.836  & 3.47 [3.03--3.91]  &  " & \nodata \Mstrut \\
c12.3  &11:50:43.603 &-28:05:13.483  & 3.36 [2.93--3.79]  &  " & \nodata \Bstrut \\
c12.4 &11:50:43.821 &-28:05:00.398  &\nodata  &  " & Blended with another object \\
\hline
c13.1  &11:50:55.879 &-28:04:25.578  & 0.81 [0.20--4.98]  &  \nodata & \nodata \Tstrut \\
c13.2  &11:50:55.428 &-28:04:19.980  & 4.16 [3.65--4.67]  &  \nodata & \nodata \Mstrut  \\
c13.3 &11:50:54.995 &-28:04:14.134  & 4.77 [4.21--5.33]  &  \nodata & \nodata \Bstrut \\
\hline
14.1  &11:50:55.751 &-28:04:04.016 & 0.77 [0.60--1.05]  & \nodata & Not used as constraint \Tstrut \\
14.2  &11:50:55.855 &-28:04:05.816  & 0.92 [0.60--1.11]  &  \nodata & " \Mstrut \\
14.3  &11:50:56.109 &-28:04:07.315  &  0.75 [0.50--0.92]  &  \nodata & " \Bstrut \\
\hline
c15.1 &11:50:50.470 &-28:03:54.703  & 0.54 [0.39--0.69] &  \nodata & Other iterations yield $z_{phot}\sim3$ \Tstrut \\
c15.2 &11:50:50.239 &-28:03:54.937   & 0.47 [0.33--0.61] &  \nodata & " \Bstrut \\
\hline
16.1  &11:50:53.949 &-28:06:17.632 & 2.97 [0.19--3.36]  &  \nodata& Not used as constraint  \Tstrut \\
16.2  &11:50:53.919 &-28:06:17.939  &\nodata  & \nodata & " \Mstrut \\
16.3  &11:50:53.588 &-28:06:19.559  & 3.37 [2.74--3.80]  &  \nodata& " \Bstrut \\
\hline
c17.1 &11:50:45.396 &-28:05:00.931  & 0.83 [0.12--3.87]  &  \nodata & \nodata \Tstrut \\
c17.2 &11:50:45.377 &-28:05:08.390  & 3.19 [0.41--3.72]  &  \nodata & Nearby parts yield  $z_{phot}\sim3.7-4$ \Bstrut \\
\hline
c18.1 &11:50:45.271 &-28:05:15.966   & 0.59 [0.12--3.46]   &  \nodata & \nodata \Tstrut \\
c18.2 &11:50:45.255 &-28:05:21.258  & 3.40 [0.69--3.83]  &  \nodata & \nodata \Bstrut \\
c17/18 & 11:50:46.578 & -28:04:23.945 & 0.71 [0.47--0.97] & \nodata & \nodata \Bstrut \\
\hline
19.1 &11:50:49.203 & -28:04:14.689 &  6.90 [6.13--8.43] &  $=$6.9 & Dropout, not detected in ACS bands \Tstrut \\ 
19.2 &11:50:52.784 & -28:04:17.603  & 6.90 [6.13--8.43]  &  " & " \Mstrut \\
19.3  &11:50:51.936 & -28:05:16.296 & 6.90 [6.13--8.54]  &  " & " \Mstrut \\
19.4  &11:50:46.185 & -28:05:44.875 & 6.94 [6.16--8.32] &  " & $"$ \Mstrut \\
c19.5  & 11:50:50.443 & -28:04:53.959& 1.35 [0.95--8.83]&  " & $"$, in BCG's light \Bstrut \\
\hline
c20.1 & 11:50:50.632 & -28:05:03.137 & 0.98 [0.07--1.71] & \nodata & In BCG light, radial image \Tstrut \\
c20.2 & 11:50:50.521 & -28:05:00.919 & 5.88 [1.60--6.55] & \nodata & $"$ \Bstrut
\enddata
\tablecomments{$\emph{Column 1:}$ ID . ``c'' stands for candidate where identification was more ambiguous, and image was not used as constraint.\\
$\emph{Columns 2 \& 3:}$ RA and DEC in J2000.0. \\
$\emph{Column 4:}$ Photometric redshift and 95\% C.L. from BPZ.\\
$\emph{Column 5:}$ Predicted and 95\% C.L. redshift by our lens model, for systems whose redshift was left to be optimized in the minimization (otherwise, a fixed value is listed).\\
$\emph{Column 6:}$ Comments.}
\end{deluxetable*}


\end{document}